\documentclass[aps,prl,superscriptaddress,reprint,floatfix]{revtex4-2}
\usepackage{graphicx}
\usepackage{dcolumn}
\usepackage{lipsum}
\usepackage{multirow}
\usepackage{amsmath}
\usepackage{newtxmath}
\usepackage{subfigure}
\usepackage{braket}
\usepackage{color}
\usepackage{booktabs}
\usepackage{colortbl}
\usepackage{rotating}
\usepackage{epstopdf}
\usepackage[normalem]{ulem}
\usepackage{natbib}
\usepackage[colorlinks,bookmarks=false,citecolor=blue,linkcolor=red,urlcolor=blue]{hyperref}
\graphicspath{{graphs/}}

\begin{document}

\title{Convert widespread paraelectric perovskite to ferroelectrics}
\author{Hongwei Wang} 
\affiliation{Department of Physics, Temple University, Philadelphia, PA 19122, USA}
\affiliation{Department of Electrical and Computer Engineering, University of Minnesota, Minneapolis, Minnesota 55455, USA}
\author{Fujie Tang} \affiliation{Department of Physics, Temple University, Philadelphia, PA 19122, USA}
\author{Massimiliano Stengel} 
\affiliation{Institut de Ciència de Materials de Barcelona (ICMAB-CSIC), Campus UAB, 08193 Bellaterra, Spain}
\affiliation{ICREA-Institució Catalana de Recerca i Estudis Avançats, 08010 Barcelona, Spain}
\author{Hongjun Xiang} 
\affiliation{Key Laboratory of Computational Physical Sciences (Ministry of Education), State Key Laboratory of Surface Physics and Department of Physics, Fudan University, Shanghai 200438, People's Republic of China}
\author{Qi An}
\thanks{$^\ast$Corresponding author. Email:qia@unr.edu}
\affiliation{Department of Chemical and Materials Engineering, University of Nevada-Reno, Reno, Nevada 89557, USA}
\author{Tony Low}
\thanks{$^\ast$Corresponding author. Email:tlow@umn.edu}
\affiliation{Department of Electrical and Computer Engineering, University of Minnesota, Minneapolis, Minnesota 55455, USA}
\author{Xifan Wu}
\thanks{$^\ast$Corresponding author. Email:xifanwu@temple.edu}
\affiliation{Department of Physics, Temple University, Philadelphia, PA 19122, USA}
\affiliation{Institute for Computational Molecular Science, Temple University, Philadelphia, PA 19122, USA}
\date{\today}

\begin{abstract}
While nature provides a plethora of perovskite materials, only a few exhibits large ferroelectricity and possibly multiferroicity. The majority of perovskite materials have the non-polar CaTiO$_3$(CTO) structure, limiting the scope of their applications. Based on effective Hamiltonian model as well as first-principles calculations, we propose a general thin-film design method to stabilize the functional BiFeO$_3$(BFO)-type structure, which is a common metastable structure in widespread CaTiO$_3$-type perovskite oxides. It is found that the improper antiferroelectricity in CTO-type perovskite and ferroelectricity in BFO-type perovskite have distinct dependences on mechanical and electric boundary conditions, both of which involve oxygen octahedral rotation and tilt.  The above difference can be used to stabilize the highly polar BFO-type structure in many CTO-type perovskite materials.
\end{abstract}

\maketitle

\par ABO$_3$ perovskite is an important family of materials that hosts many intriguing physical properties such as ferroelectricity (FE), ferromagnetism, and multiferroicity~\cite{schlom2007, zubko2011, rondinelli2011, spaldin2017}. Ferroelectricity is one of the key properties that requires a switchable polar distortion to break the inversion symmetry~\cite{rabe2007,lines2001a,tang2020p,dawber2005p,chunyizhang2022}. However, in most perovskites, polarity does not occur independently, but is accompanied or induced by primary structural distortions associated with the oxygen octahedral rotations as in the improper FE~\cite{woodward1997, benedek2013,Benedek2011i,bousquet2008i}. In general, there are two patterns of oxygen octahedral rotations in perovskites. On the one hand, perovskite such as rhombohedral BiFeO$_3$(BFO)~\cite{wang2003}, adopts an out-of-phase oxygen octahedral rotation that is compatible with a large polarization. On the other hand, perovskite such as orthorhombic CaTiO$_3$(CTO) incorporates an in-phase oxygen octahedral rotation, which gives rise to antipolar distortions with varnishing polarization~\cite{eklund2009}. Unfortunately, the CTO-type structure dominates the perovskite materials which makes the FE perovskites relatively rare in nature~\cite{benedek2013}.

\par In perovskites, in-phase and out-of-phase oxygen octahedral rotations are in strong competition energetically. Under adjustable experimental conditions, the BFO-type structures can be actually stabilized in many CTO-type perovskites~\cite{ross1989,leinenweber1991,kawamoto2014,wang2016}, and vice versa~\cite{stengel2015}. For example, the metastable BFO-type structure has been stabilized in ScFeO$_3$ under applied high pressure and high temperature~\cite{kawamoto2014}. As another example, it was theoretically proposed that CTO can be stabilized in its BFO-type phase at interfaces in superlattices~\cite{wang2016}, and the prediction was confirmed by a recent experiment~\cite{kim2020s}. 

\par Notwithstanding the significant progress achieved so far in the field, the scope of this progress is very limited. The applied hydrostatic pressure hampers the device application under ambient conditions. Artificial oxide materials based on modern thin-film technology has provided a promising route to stabilize the functional BFO-type structure~\cite{mundy2016a,eason2007p,panish1980m,dawber2005p,junquera2008f} only at interfaces~\cite{wang2016,kim2020s}, not in the bulk of the heterostructures~\cite{wang2016}. A general guideline for stabilizing the BFO-type structure over the commonly found CTO-type structure in other perovskites has eluded the community to-date~\cite{benedek2013}. More recently, it has been demonstrated~\cite{Cazorla2015} that the competing structural phases in PbTiO$_3$ and BiFeO$_3$ are very sensitive to the applied mechanical and electric boundary conditions. It indicates an alternative route to stabilize the hidden functional structure, which has not been explored so far.
 
\par To address the above issues, we propose a novel and practical design methodology, which stabilizes the functional metastable BFO-type structure in widely spread CTO-type perovskites in nature. This approach is based on proper choices of mechanical and electric boundary conditions via modern thin-film technology~\cite{dieguez2008,stengel2009}. We validate the design rule in a prototypical perovskite CdHfO$_3$~\cite{shpilevaya2004,dernier1975synthesis} by an effective Hamiltonian approach built upon the first-principles calculations, in which the epitaxial strains and electric displacement fields are considered as tunable parameters in experiments. We find that the improper mechanisms underlying the antiferroelectric (AFE) in CTO-type perovskites~\cite{Averyanova1968,Averyanova1969,shpilevaya2004} and FE in BFO-type perovskites have distinct dependences on its mechanical and electric boundary conditions. With the increased applied tensile strain, the AFE ordering via the trilinear coupling~\cite{bousquet2008i,Benedek2011i} in CTO-type structure becomes destabilized much rapidly than the FE ordering via four-linear coupling in BFO-type structure~\cite{wang2016,blok2011i}. On the other hand, the applied electric displacement field promotes the FE ordering in BFO-type structure and destabilize the AFE in CTO-type structure. Therefore, the combined epitaxial strain and applied electric field can be used to stabilize the metastable BFO-type structure in general CTO-type perovskites. We further demonstrate that such a design scheme can be broadly applied onto many CTO-type perovskites and new room-temperature artificial multiferroic materials can now be designed and synthesized.

\begin{figure}[t!]
	\setlength{\abovecaptionskip}{0.cm}
	\includegraphics[width=3.2in]{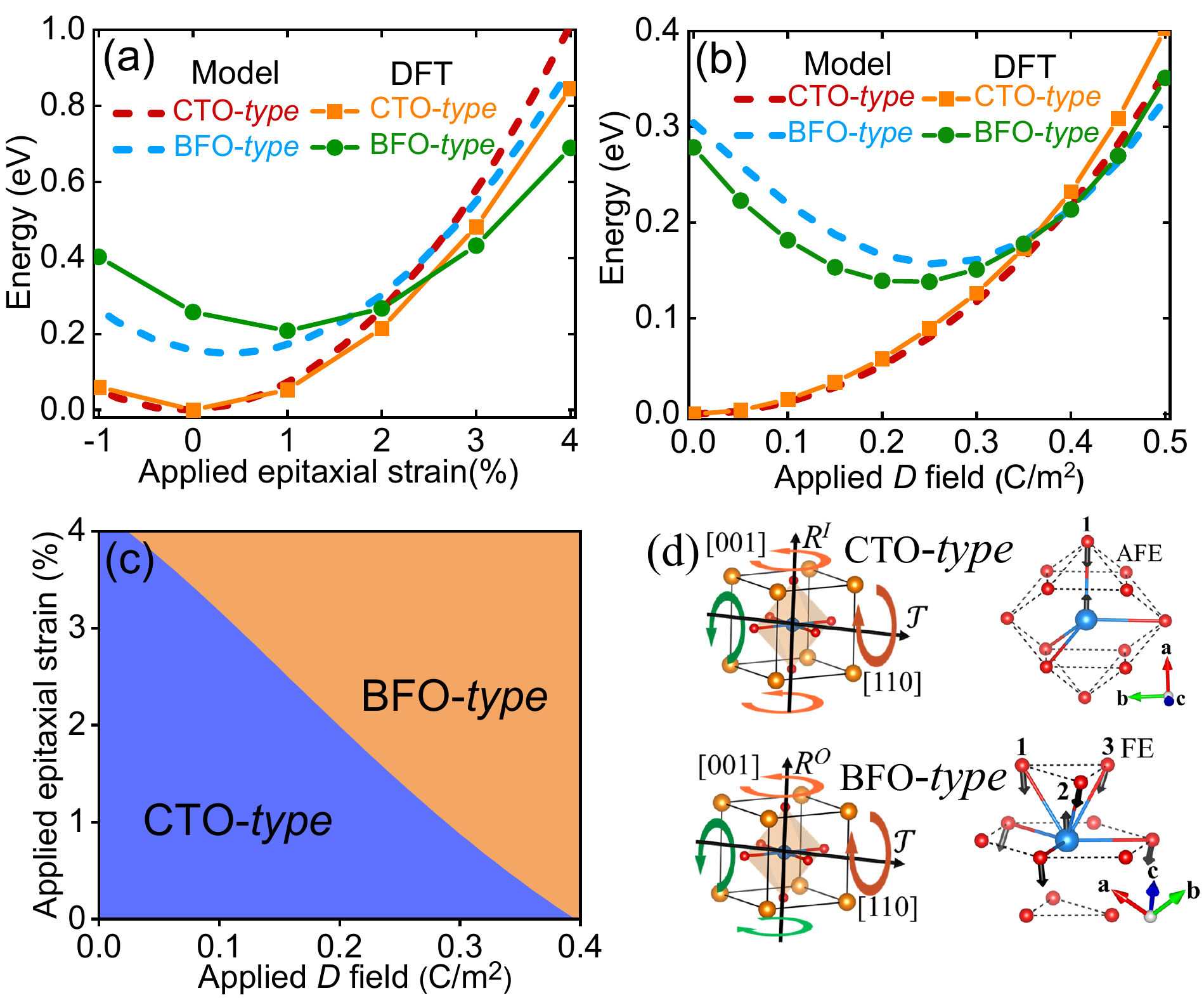}
	\caption{\label{fig:figure1}In CdHfO$_3$, the free energy of CTO-type and BFO-type structures as a function of (a) epitaxial (in-plane) strain only and zero external electric field; (b) constrained electric displacement field along z axis and zero stress in both in-plane and out-of-plane directions; (c) Phase stabilities in CdHfO$_3$ as functions of both epitaxial (in-plane) strain and constrained electric displacement field along z axis; blue (orange) denotes the CTO (BFO)-type is energetically stable. (d) Oxygen octahedral rotations and A-cation coordination environment in CTO-type and BFO-type structures. The vectors represent atom displacements in the AFE and FE modes. The oxygen atoms involved in the optimization of A-site cation coordination are labeled with 1, 2 and 3. 
		\\}
	\vspace{-3.2em}
\end{figure}

\par We start by modeling the energies of the two phases, $E_{``\rm{CTO}\text{''}}$ and $E_{``\rm{BFO}\text{''}}$ respectively obtained by density functional theory (DFT) calculations. A 20-atoms $\sqrt{2}\textbf{a}\times\sqrt{2}\textbf{a}\times2\textbf{a}$ supercell is used to accommodate FE, AFE, and oxygen octahedral rotation modes~\cite{stengel2015}, where \textbf{a} represents the pseudocubic lattice constant~\cite{glazer2011brief}. We choose CdHfO$_3$ as our prototypical perovskite, which adopts the CTO-type structure at its ground state and the BFO-type structure as the second lowest lying state with energy slightly higher ($\sim$10meV/formula, which is calculated by DFT) than the CTO-type structure. The small energy difference will facilitate the stabilization of its metastable phase by thin-film technology~\cite{lee2010,lee2011,kim2020}. Based on Landau–Devonshire theories~\cite{devonshire1951cix,dieguez2005}, the effective Hamiltonian models are built upon expanding $E_{``\rm{CTO}\text{''}}$ and $E_{``\rm{BFO}\text{''}}$ in terms of structural distortions and their intercoupling terms that are allowed by space group symmetry analysis~\cite{wang2016}. The structural distortions in the CTO-type structure[See Fig. 1(d)] are dominated by an in-phase oxygen octahedral rotation around the $[001]$ axis, an out-of-phase oxygen octahedral tilt around $[110]$ axis, and an AFE distortion mode within the AO layer, whose mode amplitudes can be denoted by $R^{\rm I}$, $\mathcal{T}$, and $\rm AFE_{xy}$ respectively. Besides the above optical phonon distortions, the strain occurring in thin-films can be described by the Voigt notation $\eta_i$ ($i=1,\ldots,6$), where $\eta_1$, $\eta_2$, and $\eta_3$ are uniaxial strains while $\eta_4$, $\eta_5$, and $\eta_6$ are shear strains. We assume a coherent growth of the perovskite film on a cubic substrate along the [001] direction, which will rule out the presence of any shear strains ($\eta_4=\eta_5=\eta_6=0$). As a result, the elastic energies will be reduced from nine to only two degrees of freedom~\cite{dieguez2005}, $\eta_1=\eta_2$=$\bar{\eta}$ and $\eta_z$. By taking both structural and strain distortions into account, we will expand the DFT energy of $E_{``\rm{CTO}\text{''}}$ in CdHfO$_3$ as a function of $\{\eta_{i}\}$ as:

\vspace{-1em}
\begin{small} 
	\begin{align}\label{eq1}
		&E_{\rm{``CTO\text{''}}}=\left[{a}_{1} {R^{\rm{I}}}^2 + {b}_{1} {R^{\rm{I}}}^4 + {a}_{2} {\mathcal{T}}^2+{b}_{2} {\mathcal{T}}^4 +{c}_{1} {R^{\rm{I}}}^2 {\mathcal{T}}^2\right]_{{E}_{\rm{``CTO\text{''}}}^{\rm{Rot}}}\nonumber \\ &+\left[\frac{1}{2}(2B_{11}\bar{\eta}^2+B_{33}{\eta}^2_z)+2B_{13}\bar{\eta}\eta_{z}+B_{12}\bar{\eta}^2\right]_{{E}_{\rm{``CTO\text{''}}}^{\rm{Elas}}} \nonumber \\
		&+\left[{a}_{3} {{\rm AFE}_{\rm{xy}}}^2+{b}_{3}{{\rm AFE}_{\rm{xy}}}^4 + {c}_{2} {R^{\rm{I}}}^2 {{\rm AFE}_{\rm{xy}}}^2+{c}_{3} {\mathcal{T}}^2 {{\rm AFE}_{\rm{xy}}}^2 \right.\nonumber \\
		&+ \left.{d}_{1} {R^{\rm{I}}}{\mathcal{T}}{{\rm AFE}_{\rm{xy}}}\right]_{{E}_{\rm{``CTO\text{''}}}^{\rm{(A)FE}}} ={E}_{\rm{``CTO\text{''}}}^{\rm{Rot}}+E_{\rm{``CTO\text{''}}}^{\rm{Elas}}+{E}_{\rm{``CTO\text{''}}}^{\rm{(A)FE}}
	\end{align}
\end{small} 
\vspace{-1.0em}

\par The distortions in the BFO-type structure[See Fig. 1(d)] can be characterized by an out-of-phase oxygen octahedral rotation around the [001] axis, an out-of-phase oxygen octahedral tilt around [110] axis, polar distortions along both in-plane ([110]) and out-of-plane ([001]) directions~\cite{wang2016}. For convenience, we will use math symbols $R^{\rm O}$, $\mathcal{T}$, and $\rm FE_{z}$ and $\rm FE_{xy}$ to denote the amplitudes of the four distortion modes, respectively. The internal energy of $E_{``\rm{BFO}\text{''}}$ of the BFO-type structure as a function of $\{\eta_{i}\}$ can be expressed as:

\vspace{-1em}
\begin{small}
	\begin{align}\label{eq2}
		&E_{\rm{``BFO\text{''}}}=\left[{\alpha}_{1}{R^{\rm{O}}}^2 + {\beta}_{1} {R^{\rm{O}}}^4 + {\alpha}_{2} {\mathcal{T}}^2 + {\beta}_{2} {\mathcal{T}}^4 + {\gamma}_{1} {R^{\rm{O}}}^2 {\mathcal{T}}^2 \right]_{{E}_{\rm{``BFO\text{''}}}^{\rm{Rot}}} \nonumber \\
		&+\left[\frac{1}{2}(2B_{11}\bar{\eta}^2+B_{33}{\eta}^2_z)+2B_{13}\bar{\eta}\eta_{z}+B_{12}\bar{\eta}^2\right]_{{E}_{\rm{``BFO\text{''}}}^{\rm{Elas}}} +\left[{\alpha}_{3}{{\rm FE}_{\rm{z}}}^2\right. \nonumber \\
		&+\left. {\beta}_{3} {{\rm FE}_{\rm{z}}}^4+{\alpha}_{4}{{\rm FE}_{\rm{xy}}}^2+ {\beta}_{4} {{\rm FE}_{\rm{xy}}}^4 +{\gamma}_{2} {{\rm FE}_{\rm{z}}}^2 {{\rm FE}_{\rm{xy}}}^2+{\gamma}_{3} {R^{\rm{O}}}^2 {{\rm FE}_{\rm{z}}}^2 \right. \nonumber \\
		&+\left.{\gamma}_{4}{R^{\rm{O}}}^2 {{\rm FE}_{\rm{xy}}}^2+{\gamma}_{5} {\mathcal{T}}^2 {{\rm FE}_{\rm{z}}}^2+{\gamma}_{6} {\mathcal{T}}^2 {{\rm FE}_{\rm{xy}}}^2 \right. \nonumber \\ 	  	  
 	    &+\left.{\delta}_{1} {R^{\rm{O}}} {\mathcal{T}} {{\rm FE}_{\rm{z}}} {{\rm FE}_{\rm{xy}}}\right]_{{E}_{\rm{``BFO\text{''}}}^{\rm{(A)FE}}}={E}_{\rm{``BFO\text{''}}}^{\rm{Rot}}+{E}_{\rm{``BFO\text{''}}}^{\rm{Elas}}+E_{\rm{``BFO\text{''}}}^{\rm{(A)FE}}
	\end{align}	
\end{small}
\vspace{-1.0em}

\par In Eqs.~\eqref{eq1} and \eqref{eq2}, the coefficients of the energy terms such as $a(\{\eta_{i}\})$  and $b(\{\eta_{i}\})$ are strain-dependent, detailed expressions are shown in Supplemental Materials(SMs). We further group all the energy terms into subcategories according to their physical effects. $E^{\rm Rot}$ refers to the energy terms contributed by the oxygen octahedral rotation and tilt as well as their couplings; $E^{\rm Elas}$ denotes the elastic energy; $E^{\rm (A)FE}$ represents the energy contributions associated with electric dipole orderings including both AFE and FE. It should be noted that the trilinear coupling term of ${d}_{1}(\{\eta_{i}\}) {R^{\rm{I}}}{\mathcal{T}}{{\rm AFE}_{\rm{xy}}}$ of Eq.~\eqref{eq1} and the four-linear coupling energy ${\delta}_{1}(\{\eta_{i}\}) {R^{\rm{O}}}{\mathcal{T}}{{\rm FE}_{\rm{z}}}{{\rm FE}_{\rm{xy}}}$ of Eq.~\eqref{eq2} play the key roles in giving rise to improper AFE and improper FE in CTO-type structure~\cite{Benedek2011i,bousquet2008i} and BFO-type structure~\cite{wang2016,blok2011i}.

\par We first explore the stabilities of both BFO-type and CTO-type structures in a CdHfO$_3$ film under tunable mechanical boundary conditions. Because of the epitaxial-strain constraint, the most stable phase cannot be determined as the structure that has the lowest unconstrained total energy. Instead, one should compare the free energies $F_{\rm{``CTO\text{''}}}(\bar{\eta})$ and $F_{\rm{``BFO\text{''}}}(\bar{\eta})$ in both structures. $F_{\rm{``CTO\text{''}}}(\bar{\eta})$ and $F_{\rm{``BFO\text{''}}}(\bar{\eta})$  should be obtained by minimizing $E_{\rm{``CTO\text{''}}}$ and $E_{\rm{``BFO\text{''}}}$ under the constraint $\eta_1=\eta_2=\bar{\eta}$, while allowing all the other degrees of freedom to be fully released, as $F_{\rm{``CTO\text{''}}}(\bar{\eta})$=$\min\limits_{\eta_1=\eta_2=\bar{\eta}}E_{\rm{``CTO\text{''}}}(\{R^{\rm{I}}\},\{\mathcal{T}\},\{{\rm{AFE}}_{xy}\},\{\eta_{z}\})$ and $F_{\rm{``BFO\text{''}}}(\bar{\eta})$=$\min\limits_{\eta_1=\eta_2=\bar{\eta}}$$E_{\rm{``BFO\text{''}}}(\{R^{\rm{O}}\},\{\mathcal{T}\},\{{\rm{FE}}_{xy}\},\{{\rm{FE}}_{z}\},\\
\{\eta_{z}\})$. The relative energy between BFO-type and CTO-type structures can be further evaluated by $\Delta F(\bar{\eta})=F_{\rm{``BFO\text{''}}}(\bar{\eta})-F_{\rm{``CTO\text{''}}}(\bar{\eta})$. If $\Delta F(\bar{\eta})$ is positive (negative), then the CTO-type (BFO-type) structure is the stable structure under this particular epitaxial strain $\bar{\eta}$.

As shown in Fig. 1(a), we display the structural phase diagram obtained by both model and direct DFT calculations as a function of epitaxial strain in CdHfO$_3$. As expected, CdHfO$_3$ bulk adopts the CTO-type structure, as evidenced by $\Delta F(\bar{\eta})>0$ when $\bar{\eta}=0$. However, $\Delta F(\bar{\eta})$ quickly drops as the applied strain increases in Fig. 1(a), and both our model and direct DFT calculations predict that the BFO-type structure is stabilized for $\bar{\eta}>2.5\%$. In order to elucidate the physical origins of the stabilized BFO-type structure under tensile epitaxial strains, we further decompose the free energies $F_{\rm{``CTO\text{''}}}(\bar{\eta})$, $F_{\rm{``BFO\text{''}}}(\bar{\eta})$, as well as $\Delta F(\bar{\eta})$ into the contribution from oxygen octahedral rotation and tilt $\Delta F^{\rm Rot}$, AFE and its coupling terms $\Delta F^{\rm (A)FE}$, as well as the elastic energy $\Delta F^{\rm Elas}$ as defined in Eqs. \eqref{eq1} and \eqref{eq2}. The resulting decompositions are shown in Figs. 2(a)-2(c).

\par Under tensile strain, the overall structural distortions associated with the octahedral octahedron and tilt are decreased in a similar extent for the CTO-type and BFO-type structures, thus their energies $F^{\rm{Rot}}_{\rm{``CTO\text{''}}}$ and $F^{\rm{Rot}}_{\rm{``BFO\text{''}}}$ increase with a similar magnitude. Due to the above, the resulting $\Delta F(\bar{\eta})^{\rm Rot}=F^{\rm{Rot}}_{\rm{``BFO\text{''}}}-F^{\rm{Rot}}_{\rm{``CTO\text{''}}}$ is roughly a positive constant value which barely contributes the stabilized BFO-type structure under tensile strain as observed in Fig. 2(a). Similarly, elastic energy does not play a decisive role in stabilizing the BFO-type structure under tensile strain. Both $F^{\rm{elas}}_{\rm{``CTO\text{''}}}(\bar{\eta})$ and $F^{\rm{FE}}_{\rm{``BFO\text{''}}}(\bar{\eta})$ show a very similar quadratic dispersion under strain, thus $\Delta F^{\rm Elas}(\bar{\eta})=F^{\rm{Elas}}_{\rm{``BFO\text{''}}}(\bar{\eta})-F^{\rm{Elas}}_{\rm{``CTO\text{''}}}(\bar{\eta})$ barely changes in Fig. 2(b).

\par Finally, we focus our attention on the energy $\Delta F(\bar{\eta})^{\rm(A)FE}$=$F^{\rm{FE}}_{\rm{``BFO\text{''}}}-F^{\rm{AFE}}_{\rm{``CTO\text{''}}}$, which decreases drastically with the epitaxial strain in Fig. 2(c) and plays a key role in stabilizing the metastable phase. In Fig. 2(c), it is evident that $\Delta F(\bar{\eta})^{\rm (A)FE}$ is dominated by the trilinear and four-linear coupling energies. The CTO-type structure rapidly becomes unstable due to the significantly increased ${d}_{1}(\{\eta_{i}\}) {R^{\rm{I}}}{\mathcal{T}}{{\rm AFE}_{\rm{xy}}}$ term in Eq.~\eqref{eq1} but the smaller change of ${\delta}_{1}(\{\eta_{i}\}) {R^{\rm{O}}}{\mathcal{T}}{{\rm FE}_{\rm{z}}}{{\rm FE}_{\rm{xy}}}$ in Eq.~\eqref{eq2}. 

\par We attribute the distinct dependences of improper AFE (CTO-type) and FE (BFO-type) on epitaxial strains to the very different capabilities in optimizing A-site coordination environments. Woodward et al. \cite{woodward1997,woodward1997octahedral} discovered that the distortions of oxygen atoms in perovskite can be attributed to the optimization of the A-site cation coordination environment, which requires that the number of short A-O interactions be maximized. The strong sensitivity of the trilinear coupling to the epitaxial strain is due to the fact that the in-plane AFE mode is directly coupled to the strain which will stretch or compress the AO bonds along roughly the same direction as schematically shown in Fig. 1(d). In contrast, the above scenario is absent in the BFO-type structure, where the improperly induced FE is along the [111] direction, whose A-site coordination is much less sensitive to the epitaxial strain applied within the AO layer as schematically plotted in Fig. 1(d).

\begin{figure}[t!]
	\setlength{\abovecaptionskip}{0.cm}
	\includegraphics[width=3.4in]{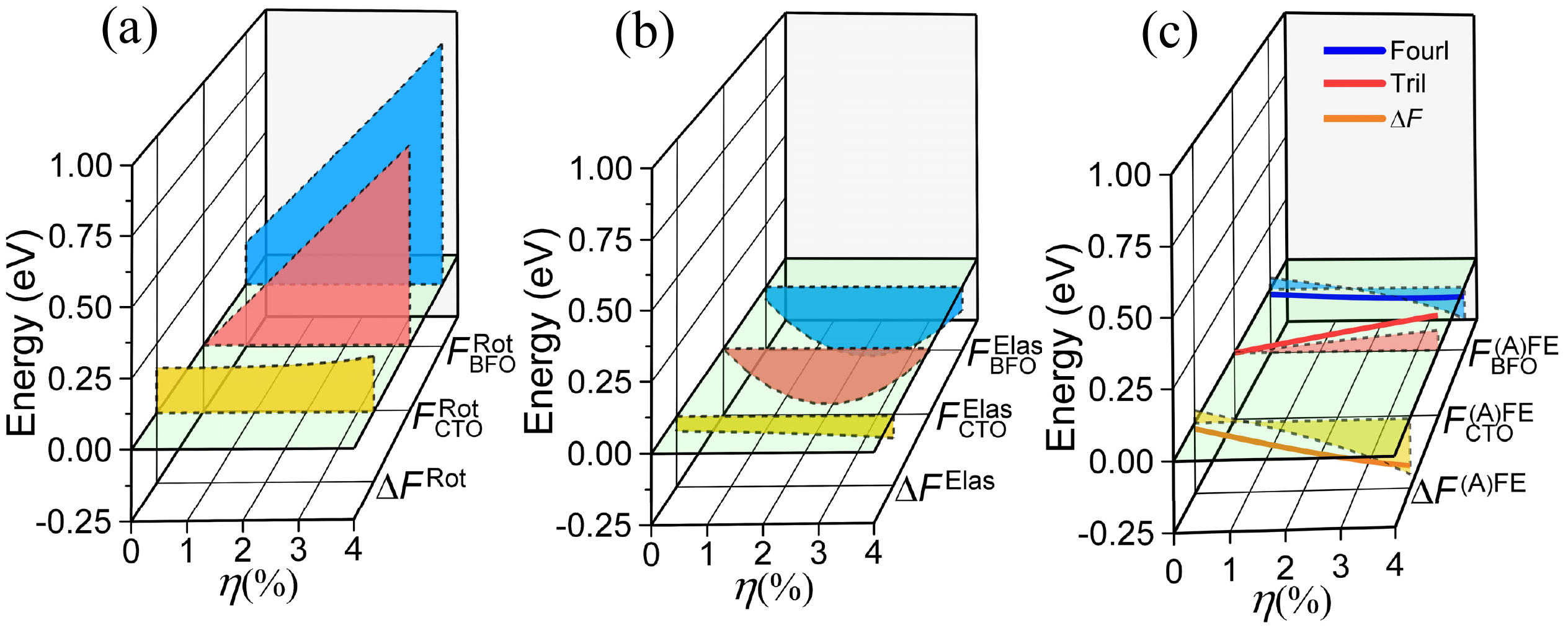}
	\caption{\label{fig:figure2}The change of free energy $\Delta F(\bar{\eta})$/unit cell (20-atom) as a function of epitaxial strain is decomposed into contributions from (a) oxygen rotation and tilt, (b) elastic energy, and (c) polar ordering energies of AFE or FE, in which the energy terms of trilinear and four-linear couplings are plotted as solid lines. The energy of unstrained CTO-type structure is set as the reference.\\}
	\vspace{-3.2em}
\end{figure}

\par Besides the epitaxial strain, the electric boundary condition offers another tuning knob to control the stabilities among various competing structures in modern thin-film technology. In order to explore the possibilities for stabilizing the BFO-type structure by the applied electric field, we will choose the electric boundary conditions by constraining electric displacement ($\emph{\textbf{D}}$) field along the thin-film growth direction [001]~\cite{stengel2009f,stengel2009}. The advantage of constrained $\emph{\textbf{D}}$-field scheme is the energy landscape becomes a single-valued function of $\emph{\textbf{D}}$-field allowing the complete probing of all stable and metastable regions. In many CTO-type perovskites, such as CdHfO$_3$, the FE mode is one of the low-lying instabilities. Therefore, the applied electric field is likely to activate this FE mode and develop its coupling terms in the modeling of $E_{``\rm{CTO}\text{''}}$. As such, the additional terms in ${E}_{\rm{``CTO\text{''}}}^{\rm{(A)FE}}$ will be added to the original Eq.\eqref{eq1}, the modified equation is shown as Eq. (S1) in SMs. Then the enthalpy functions $\rm{U}(\rm{D})$ under constrained $\emph{\textbf{D}}$-field for CTO-type and BFO-type structures can be expressed as modified formulas based on the expressions introduced by Stengel et al.~\cite{stengel2009f,stengel2009}.

\par Similarly, under the constraint electric boundary condition, we need to compare the energies of $F_{``\rm{CTO}\text{''}}(D)$ and $F_{``\rm{BFO}\text{''}}(D)$ obtained by minimizing $U(D)_{``\rm{BFO}\text{''}}$ and $U(D)_{``\rm{CTO}\text{''}}$ to determine the most stable phase at fixed $\emph{\textbf{D}}$-field. Again, the relative stability between BFO-type and CTO-type structures can be evaluated by $\Delta F(D)=F_{``\rm{BFO}\text{''}}(D)-F_{``\rm{CTO}\text{''}}(D)$. The resulting phase diagram as a function of $\emph{\textbf{D}}$-field is presented in Fig. 1(b), in which DFT results again show a reasonable agreement with our model Hamiltonian predictions. The BFO-type structure is gradually favored with the increased $\emph{\textbf{D}}$-field as indicated by the rapidly dropping value of $\Delta F(D)$. The BFO-type becomes the stable structure when the critical $\emph{\textbf{D}}$-field ($D \sim 0.4 \rm C/m^2$) is reached in CdHfO$_3$.

\par In order to unravel the microscopic origin of the stabilized BFO-type structure under $\emph{\textbf{D}}$-field, we again perform a rigorous decomposition of the free energy changes into contributions from elastic, rotations, and electric polar ordering energies as $\Delta F(D)= \Delta F^{\rm Rot}(D)+\Delta F^{\rm Elas}(D)+\Delta F^{\rm (A)FE}(D)$ as shown in Fig. 3. The elastic energy contribution $\Delta F^{\rm Elas}(D)$ to the free energy change is a rather flat dispersion in Fig. 3(b) due to the similar elastic constants of the BFO-type and CTO-type structures, which contributes little to stabilization of BFO-type structure under applied $\emph{\textbf{D}}$-field.

\par Under the applied $\emph{\textbf{D}}$-fields, FE mode develops along the [001] field direction. In typical perovskites, the polar distortion is in strong competition with octahedral rotation and tilt~\cite{vanderbilt:1998,zhong:1995}. Therefore, the octahedral rotation and tilt are gradually suppressed as a function of increased $\emph{\textbf{D}}$-field. In both CTO-type and BFO-type structures, the diminished octahedral rotation and tilt are reflected by the rapidly increased energy in Fig. 3(a), however with similar variations in both types of structures. Therefore, the energy associated with oxygen octahedral rotations and tilts $\Delta F^{\rm Rot}(D)$ does not contribute significantly to the stabilized BFO-type structure neither.

\par We next draw our attention on the free energy changes of $\Delta F^{\rm (A)FE}(D)$ associated with the FE and AFE orderings. Under applied $\emph{\textbf{D}}$-field, the CTO-type structure with improper AFE becomes more unstable because the aforementioned suppressed oxygen octahedral rotation and tilt modes make the trilinear coupling ${d}_{1}(\{\eta_{i}\}) {R^{\rm{I}}}{\mathcal{T}}{{\rm AFE}_{\rm{xy}}}$ energetically unstable. On the country, the four-linear coupling energy ${\delta}_{1}(\{\eta_{i}\}) {R^{\rm{O}}}{\mathcal{T}}{{\rm FE}_{\rm{z}}}{{\rm FE}_{\rm{xy}}}$ behind the improper FE in BFO-type structure becomes more favorable because of the promoted FE mode along the [001] $\emph{\textbf{D}}$-field direction. As a result, $\Delta F^{\rm (A)FE}(D)$ undergoes a drastic drop under $\emph{\textbf{D}}$-field as shown in Fig. 3(c), and thus plays an important role to stabilize BFO-type structure. 

\begin{figure}[t!]
	\setlength{\abovecaptionskip}{0.cm}
	\includegraphics[width=3.4in]{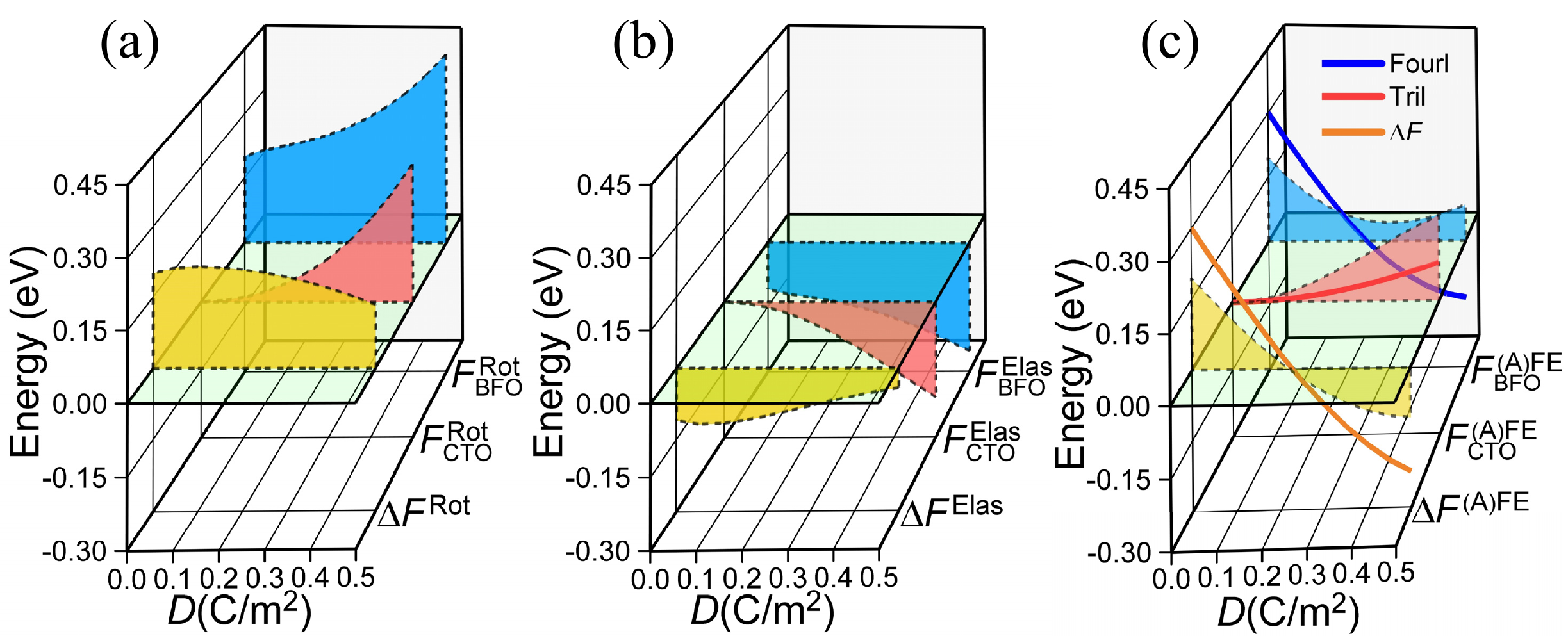}
	\caption{\label{fig:figure3}The change of free energy $\Delta F(D)$/unit cell (20-atom) as a function of electric displacement field is decomposed into contributions from (a) oxygen rotation and tilt, (b) elastic energy, and (c) polar ordering energies of AFE or FE, in which the energy terms of trilinear and four-linear couplings are plotted as solid lines. The energy of  CTO-type structure without electric field is set as the reference.\\}
	\vspace{-3.0em}
\end{figure}

\par For the thin-film or superlattices grown by epitaxial strain technique, both the applied strain and the electric field on the perovskite should be kept as small as possible in order to ensure a coherent growth and to avoid the electrical breakdown. While the above requirement might be difficult to achieve by relying on one single approach, a combined engineering method by both mechanical and electric conditions should provide a more effective way to reduce the critical fields in both directions. As shown in Fig. 1(c), we present the phase diagram predicted by model as functions of both epitaxial stain ($\bar{\eta}$) and $\emph{\textbf{D}}$-field. The blue and orange region indicate that the stable structure are the CTO-type and BFO-type, respectively. The critical epitaxial strain and $\emph{\textbf{D}}$-field to stabilize the BFO-type structure are as large as  $\bar{\eta}=2.3\%$ and $D=0.4 \rm{C/m^2}$. However, with combined boundary conditions, our preliminary studies by the model found that the critical conditions can be reduced to $\bar{\eta}=1.6\%$ and $D=0.24 \rm{C/m^2}$.

\par We then generalize the above design mechanism to the widely spread CTO-type perovskite materials in nature. We propose that there are two prerequisites. Firstly, the perovskite should have a hidden polar instability in addition to the oxygen octahedral rotation and tilt, which can be screened by the calculated phonon dispersion for at the high-temperature cubic phase. In CdHfO$_3$, the three lowest unstable phonon modes are identified to at the $\Gamma$ ($\omega = 149i$), $M$ ($\omega$ = 337i), and $R$ ($\omega$ = 327i), points of the Brillouin Zone, which give rise to the polar, in-phase oxygen octahedral rotation, and out-of-phase oxygen octahedral tilt distortions, respectively. At ground state CdHfO$_3$ adopts the CTO-type structure with the hidden polar instability, FE ordering becomes a competing structure which can be stabilized under certain mechanical and electric boundary conditions. Secondly, the BFO-type structure should be a high-lying metastable state with the small energy barrier related to ground state CTO-type structure. For CdHfO$_3$, the total DFT energy difference between the CTO-type and BFO-type structures is only $\sim$10meV/formula, which indeed facilities the stabilization of its metastable phase.

\par With the above two guidelines, we have done a survey on the databased of perovskites with CTO-type as ground state structure~\cite{jain2013commentary}. We have identified 9 perovskites that have intrinsic polar instabilities in their cubic phases. We then further employ the genetic evolutionary algorithm~\cite{xiang2010} to determine their potential metastable structures as well as their energetics and 7 of 9 perovskites are found to have BFO-type structure as the lowest-energy metastable phase. All these CTO-type perovskites potential to convert BFO-type ferroelectrics are shown in the Table SVII.

\par We first stabilize the BFO-type structure by gradually increasing the magnitudes of tensile strains. As presented in Table SVII, all the seven perovskites can be stabilized into BFO-type structures if the applied tensile strain is large enough. We then proceed to apply the proper electric boundary conditions to stabilize the BFO-type structure. In particular, we carry out constrained  $\emph{\textbf{D}}$-field DFT calculations for CdSnO$_3$ and NaMnF$_3$, which are presented in Figs. S2(a) and S2(b). Both CdSnO$_3$ and NaMnF$_3$ can be stabilized to be BFO-type structure with $D=0.25 \rm{C/m^2}$ and $D=0.17 \rm{C/m^2}$ respectively. By combining the mechanical and electric boundary conditions, the critical strain and $D$-field required to stabilize the BFO-type structure can be both further optimized. Our first-principles calculations show that as large as $D=0.50 \rm{C/m^2}$ and $D=0.50 \rm{C/m^2}$ are required to stabilize BFO-type structure in CaHfO$_3$ and CaZrO$_3$ by using applied $\emph{\textbf{D}}$-field only, which
are presented in Figs. S3(a) and S3(b). However, the critical $D$-fields are largely reduced to $D=0.28 \rm{C/m^2}$ and $D=0.25 \rm{C/m^2}$ respectively for CaHfO$_3$ and CaZrO$_3$ under 2\% epitaxial stain.

\par In summary, a number of  BFO-type ferroelectric and multiferroic perovskite materials have been successfully designed through epitaxial strain and $\emph{\textbf{D}}$-field boundary conditions.
These stabilized single-phase  BFO-type perovskites can be used for engineering applications under ambient condition, which is more practical compared to other strategies such as 
high pressure, chemical substitution, and interfacial engineering. Moreover, the CTO-type LuFeO$_{3}$ possesses a high-temperature antiferromagnetic order~\cite{chowdhury2014}, therefore, the stabilized high-polar LuFeO$_{3}$ may be another multiferroic material like BFO. In the future, the studies of the phase stability by the DFT-based effective Hamiltonian approach in current work can be systematically improved by the inclusion of higher-order invariants as well as the gradient invariants in the calculation of the free energy.

\vspace{-2.0em}
\section{Acknowledgments}
\vspace{-1.0em}
\par This work was supported by National Science Foundation through Awards DMR-2053195 (X.W.). H.W. was partially supported as part of the Center for the Computational Design of Functional Layered Materials, Department of Energy, Office of Science, Basic Energy Sciences, under Grant No. DE-SC0012575. T.L. acknowledge funding support from NSF/DMREF under Grant Agreement No. 1921629. Q.A. was supported by the American Chemical Society Petroleum Research Fund (PRF\# 58754-DNI6). The computational work used resources of the National Energy Research Scientific Computing Center (NERSC), a U.S. Department of Energy Office of Science User Facility operated under Contract No. DE-AC02-05CH11231.
\vspace{-1em}

\bibliography{R3c}

\end{document}